\documentstyle[sprocl,epsfig]{article}

\newcommand{\MeV}{{\rm MeV}}
\newcommand{\GeV}{{\rm GeV}}
\newcommand{\fb}{{\rm fb}}

\begin{document}

\title{Width effects in slepton production
  {\boldmath $e^+e^- \to \tilde{\mu}^+_R \, \tilde{\mu}^-_R$}
  \footnote{Contribution to Workshop {\em `Physics at TeV Colliders'}, 
      Les Houches, France, 8 - 18 June 1999} }
\author{ Hans-Ulrich Martyn }
\address{ I. Physikalisches Institut, RWTH Aachen, Germany }

\maketitle
\abstracts{
  A case study will be presented to investigate width effects
  in the precise determination of slepton masses at the 
  $e^+e^-$ {\sc Tesla} Linear Collider. }

\section{Introduction}

If supersymmetry will be discovered in nature a precise measurement of the 
particle spectrum will be very important in order to determine the underlying 
theory.
The potential of the proposed {\sc Tesla} Linear Collider~\cite{cdr} with its 
high luminosity and polarisation of both $e^\pm$ beams will allow to obtain 
particle masses with an accuracy of $10^{-3}$ or better~\cite{lcphysics}.
At such a precision width effects of primary and secondary particles
may become non-negligible.

\begin{figure}[hbt]
\centering
\epsfig{file=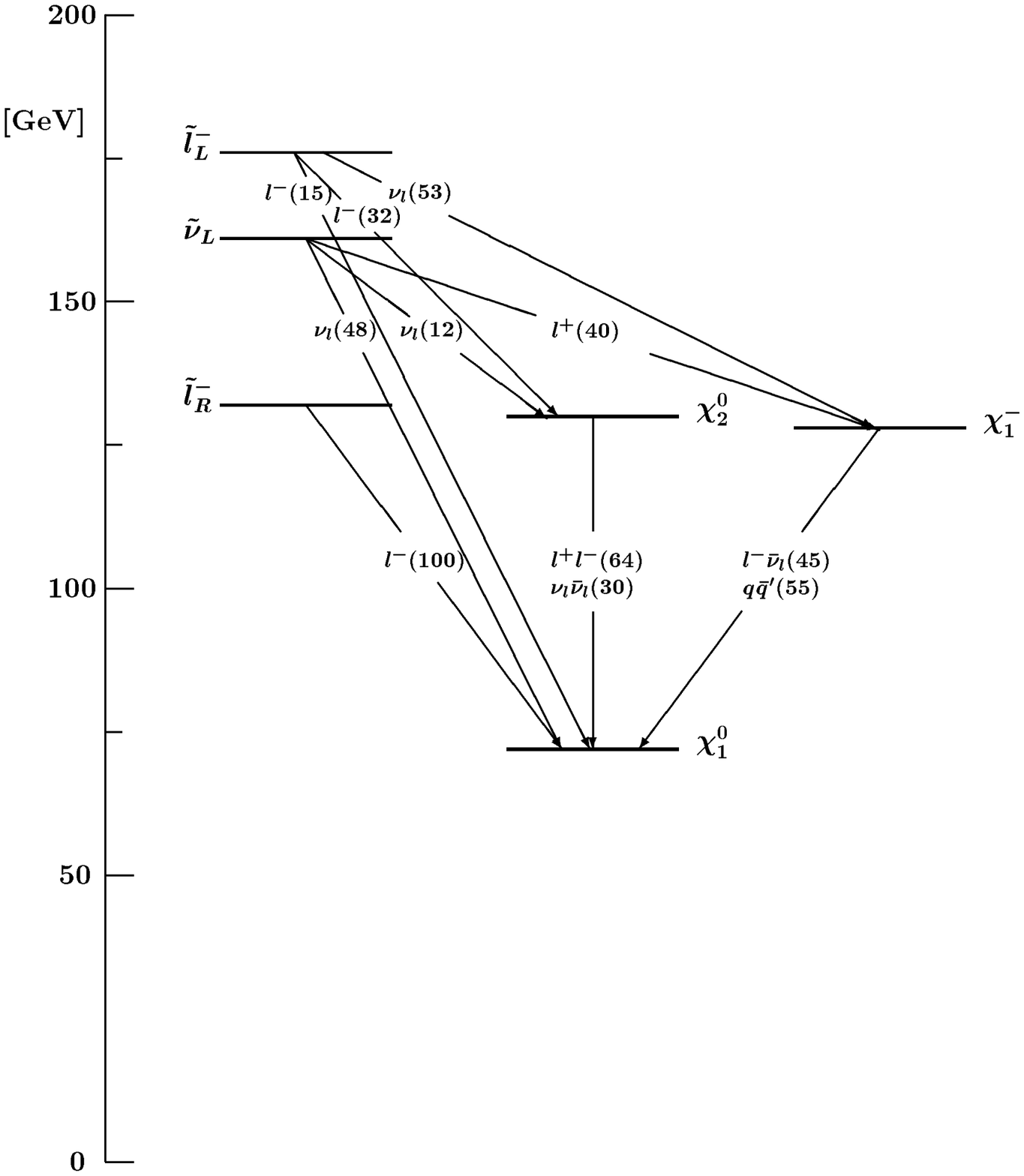,%
  bbllx=0pt,bblly=340pt,bburx=570pt,bbury=680pt,clip=,%
  height=7.5cm}
\caption{Mass spectrum and decay modes of sleptons and light gauginos
\label{fig:spectrum}}
\end{figure}

The present case study is based on a particular
$R$-parity conserving m{\sc Sugra} scenario,
also investigated in the {\sc Ecfa/Desy} Study~\cite{ecfadesy},
with parameters
$  m_0 = 100~\GeV, \ m_{1/2} = 200~\GeV, \ A_0 = 0~\GeV, \
  \tan\beta = 3$ and ${\rm sgn}(\mu) > 0 $.
The particle spectrum is shown in fig.~\ref{fig:spectrum}.
Typical decay widths of the scalar leptons are expected to be 
$\Gamma \sim 0.3 - 0.5~\GeV$, 
while the widths of the light gauginos,
decaying into 3-body final states, are (experimentally) negligible.

This note presents, as an example, a simulation of right scalar muon
production 
\begin{eqnarray}
  e^-_R e^+_L & \to & \tilde{\mu}_R^- \, \tilde{\mu}_R ^+\ ,\\
              & \to & \mu^- \chi^0_1 \,\, \mu^+ \chi^0_1 \ . \nonumber
\end{eqnarray}
The analysis is based on the methods and techniques described in a
comprehensive study of the same {\sc Susy} spectrum~\cite{mb}.
The detector concept, acceptances and resolutions are taken from the
{\sc Tesla} Conceptual Design Report~\cite{cdr}.
Events are generated with the Monte Carlo program 
{\sc Pythia}~6.115~\cite{pythia}, 
which includes the width of supersymmetric particles as well as
QED radiation and beamstrahlung~\cite{circe}.
It is assumed that both beams are polarised, right-handed electrons
to a degree of ${\cal P}_{e^-_R} = 0.80$ and left-handed positrons by
${\cal P}_{e^+_L} = 0.60$.
A proper choice of polarisations increases the cross section by a factor of 
$\sim 3$ and reduces the background substantially, e.g. by more than an 
order of magnitude for Standard Model processes.

\section{Mass determinations}

Scalar muons $\tilde{\mu}_R$ are produced in pairs via $s$ channel $\gamma$ 
and $Z$ exchange and decay into an ordinary muon and a stable neutralino 
$\chi_1^0$ (LSP), which escapes detection.
The experimental signatures are two acoplanar muons in final state with 
large missing energy and nothing else in the detector. 
Simple selection criteria~\cite{cdr} 
(essentially cuts on acollinearity angle and missing energy)
suppress background from
$W^+W^-$ pairs and cascade decays of higher mass {\sc Susy} particles
and result in detection efficiencies around $\sim 70\%$.
%The kinematics of the decay chain allow to identify and to reconstruct the
%masses of the primary and secondary sparticles.

Two methods to determine the mass of $\tilde{\mu}_R$ will be discussed:
(i) a threshold scan of the pair production cross section, and
(ii) a measurement of the energy spectrum of the decay muons,
which simultaneously constrains the mass of the primary smuon
and the secondary neutralino.
The particle mass parameters given by the chosen {\sc Susy} model
are $m_{\tilde{\mu}_R} = 132.0~\GeV$,
$\Gamma{\tilde{\mu}_R} = 0.310~\GeV$ and $m_{\chi^0_1} = 71.9~\GeV$.

\subsection{Threshold scan} 

\begin{figure}[htb]
  \centering \vspace{-.3cm}
  \epsfig{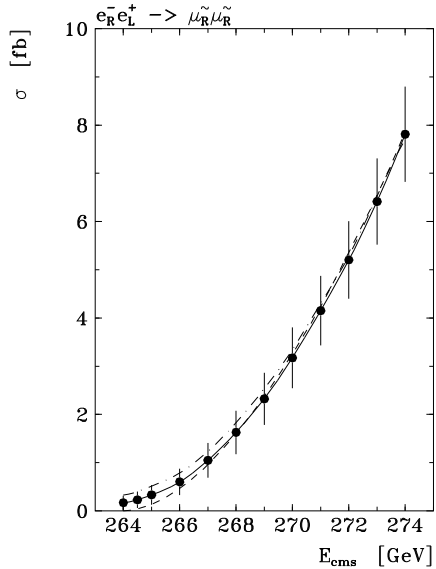}
  \vspace{-.2cm}
  \caption{ Observable cross section near threshold of the reaction
    $e^-_R e^+_L \rightarrow \tilde{\mu}_R \tilde{\mu}_R$
    including QED radiation and Beamstrahlung. 
    Curves assume a mass $m_{\tilde{\mu}_R} = 132.0~\GeV$ and width
    $\Gamma_{\tilde{\mu}_R}$ of $310~\MeV$ (full curve),
    $0~\MeV$ (dashed curve) and $620~\MeV$ (dashed-dotted curve).
    Measurements correspond to ${\cal L} = 10~\fb^{-1}$ per point.}
  \label{fig:scan}
\end{figure}

Cross section measurements close to production threshold are relatively
simple.
One essentially counts additional events with a specific signature, here
two oppositely charged, almost monoenergetic muons,
over a smooth background.
The cross section for slepton pair production rises as
$\sigma \propto \beta^3$, where 
$\beta = \sqrt{1 - 4\,m_{\tilde{\mu}_R}^2/s}$ is the velocity related to the 
$\tilde{\mu}_R$ mass.
The excitation curve as a function of the cms energy, including effects due
to QED initial state radiation and beamstrahlung,
is shown in figure~\ref{fig:scan}.
The sensitivity to the width $\Gamma_{\tilde{\mu}_R}$ is most pronounced 
close to the kinematic production limit and diminishes with inreasing energy.
A larger width `softens' the rise of the cross section with energy.
Fits to various mass and/or width hypotheses are performed by simulating
measurements with a total integrated
luminosity of $100~\fb^{-1}$ distributed over 10 
equidistant points around $\sqrt{s} = 264 - 274~\GeV$.
The data may be collected within a few months of {\sc Tesla} operation.

Taking the width from the model prediction, $\Gamma_{\tilde{\mu}_R} = 310~\MeV$,
a fit to the threshold curve gives a statistical accuracy 
of $\delta m_{\tilde{\mu}_R} = \pm 90~\MeV$
for the smuon mass.
This error is considerably smaller than the expected width.
A two-parameter fit yields
$m_{\tilde{\mu}_R} = 132.002~^{+0.170}_{-0.130}~\GeV$ and
$\Gamma_{\tilde{\mu}_R} = 311~^{+560}_{-225}~\MeV$. 
However, both parameters are highly correlated with a correlation coefficient 
of $0.95$. 
Finally, if one may fix the $\tilde{\mu}_R$ mass from another measurement, the
width can be determined to $\delta\Gamma_{\tilde{\mu}_R} = \pm 190~\MeV$.
It should be noted that the scan procedure and choice of energy 
measurement points is not optimised. 
Possibilities to reduce the correlations should be studied.

\subsection{Energy spectrum of $\mu^\pm$}

For energies far above threshold,  the kinematics of the decay chain 
of reaction (1) allows to identify and to reconstruct the
masses of the primary and secondary sparticles.
The isotropic decay of the scalar muon leads to a flat energy
spectrum of the observed final $\mu^\pm$ in the laboratory frame.
The endpoints of the energy distribution are related to the masses of
$\tilde{\mu}_R$ and $\chi^0_1$ via
\begin{eqnarray}
  E_{max,min} & = & \frac{m_{\tilde{\mu}}}{2}
    \left ( 1 - \frac{m_{\chi^0}^2}{m_{\tilde{\mu}}^2} \right )
    \gamma \, (1  \pm  \beta)  \  .
\end{eqnarray}

\begin{figure}[htb]
  \begin{center}
  \hspace{-.5cm}
  \epsfig{file=fig3a.eps,%
    bbllx=20pt,bblly=40pt,bburx=470pt,bbury=740pt,clip=,%
    angle=90,height=6cm}
  \epsfig{file=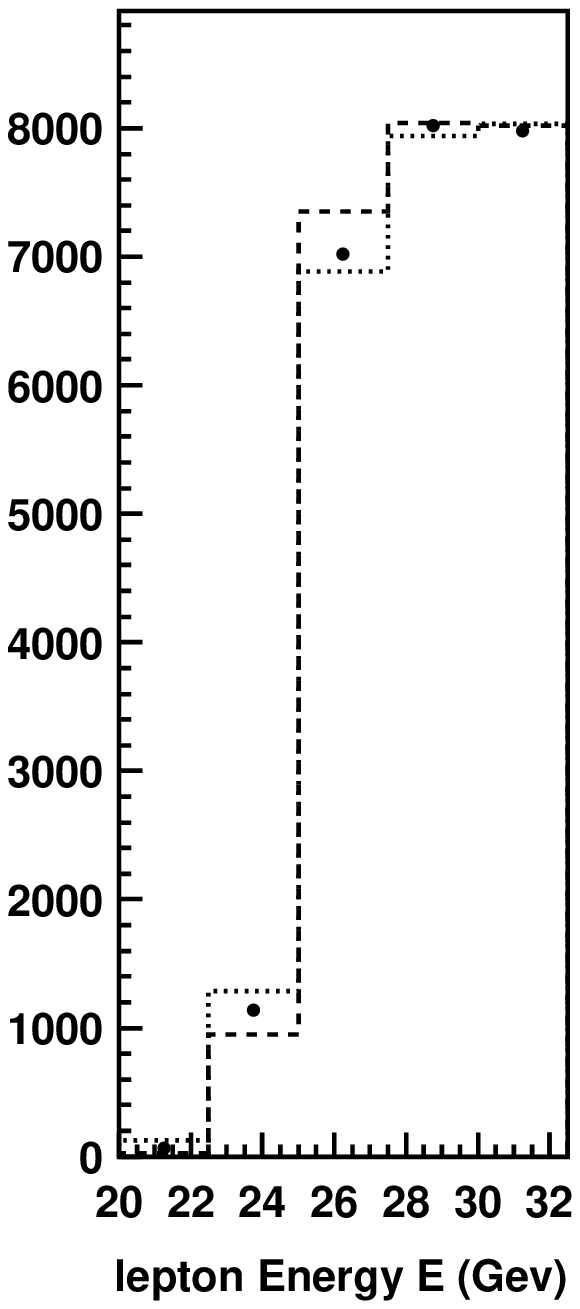,%
    bbllx=0pt,bblly=0pt,bburx=230pt,bbury=390pt,clip=,%
    angle=0,height=6cm}
  \end{center}
  \caption{{\bf Left:}
    Energy spectrum of di-muon events from the reaction
    $e^-_R e^+_L \to \tilde{\mu}_R \tilde{\mu}_R$ 
    and the background $e^-_R e^+_L \to \chi^0_1 \chi^0_2$
    at $\sqrt{s}=320~\GeV$ assuming ${\cal L} = 160~\fb^{-1}$.
    {\bf Right:} 
    Lower endpoint region of the $\mu$ energy spectrum illustrating 
    the effect of various widths $\Gamma_{\tilde{\mu}_R}$ of 
    $310~\MeV$ (dots), $0~\MeV$ (dashed) and $620~\MeV$ (dotted)
    using the tenfold luminosity.}
  \label{fig:sleptons}
\end{figure}

In practice the sharp edges of the energy spectrum will be smeared by effects
due to detector resolution, selection criteria and in particular
initial state radiation and beamstrahlung.
The results of a simulation at $\sqrt{s} = 320~\GeV$ assuming an integrated
luminosity of $160~\fb^{-1}$ are shown in figure~\ref{fig:sleptons}.
One observes a clear signal from $\tilde{\mu}_R$ pair production above a
small background of cascade decays $\chi^0_2 \to \mu^+ \mu^- \chi^0_1$
from the reaction $e^-_R e^+_L \to \chi^0_2 \chi^0_1$. 
Contamination from chargino or $W$ pair production is completely negligible.

A two-parameter fit to the $\mu$ energy spectrum yields masses of
$m_{\tilde{\mu}_R} = 132.0 \pm 0.3~\GeV$ and
$m_{\chi^0_1}      =  71.9 \pm 0.2~\GeV$.
The statistical accuracy is of the same size as the expected width of the
scalar muon.
Choosing a different width $\Gamma_{\tilde{\mu}_R}$ in the simulation modifies
essentially the $\mu$ energy spectrum at the low endpoint and has little impact
at higher energies. 
This is illustrated in figure~\ref{fig:sleptons}, right part, 
which compares the lower part of the spectrum with the predictions of width 
zero and twice the expected value.
The sharp rise is getting smeared out with increasing width.
With the anticipated luminosity of $160~\fb^{-1}$ it may be feasible to 
distinguish these cases.

\subsection{Production of other sparticles}

It should be noted that estimates on the sensitivity of width effects
in other slepton production channels can be obtained from the above results
by scaling the cross section and taking the branching ratios into final 
states into account. 
Thus one expects e.g. a gain by a factor of $\sim$2 for selectron
$\tilde{e}_R$ and sneutrino $\tilde{\nu}_e$ pair production.
For the higher mass chargino $\chi^\pm_2$ and neutralinos $\chi^0_{3,\,4}$
mass resolutions of $\sim 0.25 - 0.50~\GeV$ may be obtained from threshold
scans~\cite{mb}, where the cross sections rise as $\sigma \propto \beta$.
The corresponding widths are expected to be $\sim 2 - 5~\GeV$ 
(two-body decays in a gauge boson and gaugino) and have certainly to be
considered.

\section{Conclusions}

The high luminosity of {\sc Tesla} allows to study the production
and decays of the accessible {\sc Susy} particle spectrum. 
Polarisation of both $e^-$ and $e^+$ beams is very important to
optimise the signal and suppress backgrounds.
A simulation of slepton production $e^+e^- \to \tilde{\mu}_R \tilde{\mu}_R$
shows that for precision mass measurements with an accuracy of 
${\cal O}(100~\MeV)$ the widths of the primary particles have to be taken 
into account. 
Finally, it is worth noting that the anticipated mass resolutions 
from threshold scans or lepton energy spectra
can only be obtained if beamstrahlung effects are well under control.

\end{document}